\documentclass[sigconf]{acmart}

\AtBeginDocument{%
  }

\copyrightyear{2026}
\acmYear{2026}
\setcopyright{cc}
\setcctype{by}
\acmConference[CHI '26]{Proceedings of the 2026 CHI Conference on Human Factors in Computing Systems}{April 13--17, 2026}{Barcelona, Spain}
\acmBooktitle{Proceedings of the 2026 CHI Conference on Human Factors in Computing Systems (CHI '26), April 13--17, 2026, Barcelona, Spain}
\acmPrice{}
\acmDOI{10.1145/3772318.3790805}
\acmISBN{979-8-4007-2278-3/2026/04}

\begin{document}

\title{BRIDGE: Borderless Reconfiguration for Inclusive and Diverse Gameplay Experience via Embodiment Transformation}

\author{Hayato Saiki}
\email{saiki@ai.iit.tsukuba.ac.jp}
\orcid{0009-0005-8374-9483}
\affiliation{%
  \institution{Artificial Intelligence Laboratory}
  \institution{University of Tsukuba}
  \city{Tsukuba}
  \country{Japan}
}

\author{Chunggi Lee}
\email{chunggi_lee@g.harvard.edu}
\orcid{0000-0002-6164-2563}
\affiliation{%
  \institution{Visual Computing Group}
  \institution{Harvard University}
  \city{Cambridge}
  \state{Massachusetts}
  \country{USA}
}

\author{Hikari Takahashi}
\email{s2530479@u.tsukuba.ac.jp}
\orcid{0009-0005-2355-7711}
\affiliation{%
  \institution{Graduate School of Comprehensive Human Sciences}
  \institution{University of Tsukuba}
  \city{Tsukuba}
  \country{Japan}
}

\author{Tica Lin}
\email{mlin@g.harvard.edu}
\orcid{0000-0002-2860-0871}
\affiliation{%
  \institution{Dolby Laboratories}
  \city{Atlanta}
  \state{Georgia}
  \country{USA}
}

\author{Hidetada Kishi}
\email{h.kishi@mejiro.ac.jp}
\orcid{0009-0003-4113-5737}
\affiliation{%
  \institution{Mejiro University}
  \city{Shinjuku-ku}
  \state{Tokyo}
  \country{Japan}
}

\author{Kaori Tachibana}
\email{tachibana@ipu.ac.jp}
\orcid{0000-0002-1669-0689}
\affiliation{%
  \institution{Ibaraki Prefectural University of Health Sciences}
  \city{Ibaraki}
  \country{Japan}
}

\author{Yasuhiro Suzuki}
\email{yasuhiro@ai.iit.tsukuba.ac.jp}
\orcid{0000-0002-5141-6941}
\affiliation{%
  \institution{Institute of Systems and Information Engineering}
  \institution{University of Tsukuba}
  \city{Tsukuba}
  \country{Japan}
}

\author{Hanspeter Pfister}
\email{pfister@seas.harvard.edu}
\orcid{0000-0002-3620-2582}
\affiliation{%
  \institution{Visual Computing Group}
  \institution{Harvard University}
  \city{Cambridge}
  \state{Massachusetts}
  \country{USA}
}

\author{Kenji Suzuki}
\email{kenji@ieee.org}
\orcid{0000-0003-1736-5404}
\affiliation{%
  \institution{Institute of Systems and Information Engineering}
  \institution{University of Tsukuba}
  \city{Tsukuba}
  \country{Japan}
}


\begin{abstract}
Training resources for parasports are limited, reducing opportunities for athletes and coaches to engage with sport-specific movements and tactical coordination. To address this gap, we developed BRIDGE, a system that integrates a reconstruction pipeline, which detects and tracks players from broadcast video to generate 3D play sequences, with an embodiment-aware visualization framework that decomposes head, trunk, and wheelchair base orientations to represent attention, intent, and mobility. We evaluated BRIDGE in two controlled studies with 20 participants (10 national wheelchair basketball team players and 10 amateur players). The results showed that BRIDGE significantly enhanced the perceived naturalness of player postures and made tactical intentions easier to understand. In addition, it supported functional classification by realistically conveying players’ capabilities, which in turn improved participants’ sense of self-efficacy. This work advances inclusive sports learning and accessible coaching practices, contributing to more equitable access to tactical resources in parasports.
\end{abstract}


\begin{teaserfigure}
  \includegraphics[width=\textwidth]{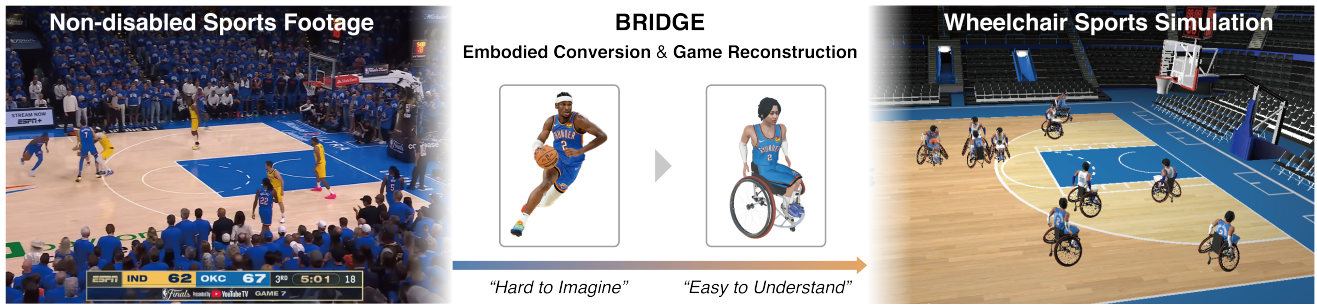}
  \caption{An overview of the BRIDGE system. Given non-disabled sports footage as input, the system reconstructs the game and maps player embodiments to wheelchair-specific constraints. This bridging process generates a wheelchair sports scene in 3D space, enabling wheelchair athletes to better understand and engage with the gameplay. Images include still frames from NBA broadcast footage and a player photograph used for research illustration purposes. © NBA / Broadcast rights holders. © Joshua Gateley / Getty Images.}
  \Description{Overview of the BRIDGE system showing how non-disabled basketball broadcast footage is transformed into a 3D wheelchair basketball simulation through embodiment-aware mapping.}
  \label{teaser}
\end{teaserfigure}


\maketitle

\section{INTRODUCTION}
Video analysis has become a cornerstone of tactical learning in modern sports. In disciplines such as basketball, soccer, and American football, players and coaches increasingly rely on video to analyze strategies, anticipate opponents’ moves, and refine decision-making~\cite{naik2022comprehensive}. Professional leagues and training programs now provide extensive video resources, enabling athletes at all levels to study advanced tactics and improve their skills.

However, this abundance is not equally distributed across all sports. In parasports, the limited player population and media exposure result in a severe scarcity of video resources~\cite{rodriguez2022sport, khurana2021beyond}. As observed in our formative study, many wheelchair basketball players and coaches rely on non-disabled sports footage for training and analysis (Section~\ref{sec:formative}). While these materials are tactically informative, the differences in physical embodiment pose a significant barrier for para-athletes. In wheelchair-based sports, athletes propel themselves using wheelchairs rather than relying on bipedal locomotion, fundamentally reshaping feasible movement patterns, action repertoires, and orientation cues~\cite{hanks2018muscle, churton2013constraints}. For example, turning requires coordinated wheel maneuvers rather than a single step, defensive coverage depends on wheelchair acceleration limits, and players’ head orientation is not always aligned with their actual direction of movement. 
As a result, tactical elements from standing or running sports, such as rapid accelerations, sharp directional changes, or body orientation signals, cannot be directly transferred or meaningfully interpreted in the wheelchair context. Furthermore, in parasports, athletes with diverse physical abilities compete under rules that account for functional differences. Interpreting non-disabled sports footage while considering these class-based constraints imposes substantial cognitive demands on both players and coaches~\cite{IWBF_classification_basics}.
Beyond these biomechanical mismatches, reliance on sports footage featuring non-disabled players also risks undermining athletes’ self-efficacy and sense of affirmation, as their lived sporting experiences are not adequately represented. Prior work has emphasized that in parasport contexts, the meaning and value of sport must be reinterpreted and extended to align with the embodied realities of athletes with disabilities~\cite{strobel2025hci}. Accordingly, when leveraging sports videos featuring different physical embodiments, there is a critical need to adapt and extend such materials for parasport use.

To address these challenges, our research explores methods for transforming sports video to support tactical learning across embodiments. We focus specifically on wheelchair basketball, where athletes must reinterpret stand-up basketball strategies under distinct physical constraints. We conducted a formative study with eight national team players and two coaches to identify the main difficulties athletes face in interpreting tactics. The study revealed that participants struggled to adapt stand-up basketball footage to wheelchair basketball, citing three main challenges: \textbf{1) imagining the movements in a wheelchair}, \textbf{2) differences in body orientation}, and \textbf{3) large variations in execution across functional classes}.
Informed by the study, we propose \textbf{BRIDGE}, a system that reconstructs stand-up basketball footage into the tactical and embodied context of wheelchair basketball. BRIDGE integrates a four-stage 3D reconstruction process, consisting of \textit{player and ball detection and tracking, court coordinate mapping}, \textit{ball possession}, and \textit{action state estimation}, and \textit{trajectory smoothing}. It also includes an \textbf{embodiment-aware conversion framework} that decomposes head, trunk, and wheelchair base orientations. This hierarchical mapping enables realistic representation of functional classifications, embodied action, and feasible mobility. By transforming broadcast basketball footage into a virtual wheelchair basketball simulation (Figure~\ref{teaser}), BRIDGE provides a novel conversion system that supports players and coaches in imagining execution, interpreting posture and orientation, and adapting tactics across functional classes.

To validate the effectiveness of our approach, we conducted two experiments involving 10 national team players and 10 non-elite players. These studies systematically evaluated how embodiment-aware orientation mapping and videos transformed by BRIDGE influence functional limitation inference, tactical understanding, perceived realism, and self-efficacy. The results showed that orientation mapping not only enhanced the perceived naturalness of postures but also enabled appropriate representation of functional differences. Furthermore, the wheelchair-converted videos were found to \textbf{strengthen both self-efficacy and learning motivation}, thereby lowering the sense of alienation and entry barriers that have traditionally hindered inclusive sports learning. Together, these findings demonstrate that our system provides an effective means of bridging the contexts of stand-up and wheelchair basketball.
In summary, our contributions are as follows:

\textbf{(1)} A design framework for embodiment-aware orientation mapping, decomposing and remapping head, trunk, and wheelchair base orientations to visually represent attention, action, and mobility.  

\textbf{(2)} The design and implementation of a novel conversion technique that reconstructs stand-up basketball footage into wheelchair basketball contexts, bridging tactical learning across embodiments.  

\textbf{(3)} An empirical evaluation with 20 participants, including 10 national wheelchair basketball players, demonstrating how orientation mapping and cross-embodiment transformation affect tactical understanding, functional limitation inference, perceived naturalness, and self-efficacy.

\textbf{(4)} Design implications for inclusive sports learning, highlighting how cross-embodiment conversion not only extends tactical resources to underrepresented athletes but also fosters equitable participation and motivation across diverse learner groups.
\section{RELATED WORK}

\subsection{Inclusive Learning in Parasports}

Research on parasports has gained increasing attention in recent years, with several reviews highlighting its significance~\cite{strobel2025hci, motahar2025understanding, khurana2021beyond}. Compared to non-disabled sports, however, system development and empirical studies directly targeting parasport athletes remain limited. Still, pioneering technology-driven initiatives have emerged. For instance, \textit{VolleyNaut}, a VR training system for sitting volleyball, enables athletes with disabilities to practice defensive skills in immersive environments and shows potential for performance enhancement~\cite{gong2024volleynaut}. In a similar vein, \textit{Paralympic VR Game} by Macedo et al.\ allows users to experience wheelchair basketball from an athlete’s perspective, combining 3D graphics and 360-degree video to enhance empathy toward Paralympic sports and evaluate immersion and motion sickness~\cite{macedo2019paralympic}. Likewise, \textit{ARSports}, a wearable AR system, leverages computer vision models to extend basketball and tennis play in near real-time for people with visual impairments~\cite{lee2024towards}.
Accessibility efforts have also targeted spectator experiences. NTT’s \textit{Kirari!} applies spatial audio to allow visually impaired audiences to engage with games through sound alone~\cite{Miyakawa2022Kirari}. Furthermore, insights from rehabilitation and accessibility studies have been applied in sports contexts, such as a wearable device that guides visually impaired runners via tactile cues~\cite{machado2021pro} and a haptic feedback system that supports throwing actions in the Finnish game Mölkky~\cite{kobayashi2022accessibility}.
Sensing-based approaches are also emerging, providing a foundation for future developments~\cite{kelly2024wheelskills}. For example, IMU- and accelerometer-based methods have been used to analyze wheelchair propulsion in real-time for performance evaluation and injury prevention~\cite{weizman2024application}. In addition, \textit{WheelPoser}, which estimates joint postures of wheelchair users with only four IMUs, demonstrates the potential of motion estimation not only for assessment but also for interactive applications~\cite{li2024wheelposer}.

While VR, AR, and sensing environments are gradually being established in parasports, systematic knowledge remains insufficient, and little research has sought to bridge the environmental gap between parasports and non-disabled sports. Against this backdrop, our study investigates whether the information-rich environments of non-disabled sports can be applied to parasports, offering novel design insights for HCI that advance inclusive learning environments.

\subsection{Embodiment Reinterpretation}
In HCI, embodiment has been understood as a process in which action, perception, and meaning are interconnected through the body~\cite{marshall2013theories, serim2024revisiting}. This perspective provides a theoretical basis for phenomena such as \textit{Homuncular Flexibility}—the ability to reconfigure bodily form—and the \textit{Proteus Effect}, whereby avatar characteristics influence self-perception and behavior~\cite{won2015homuncular, yee2007proteus}. In other words, humans do not treat the body as fixed, but reinterpret it flexibly in relation to context and external representations.  

Building on this framework, empirical studies have shown that users can project their movements onto non-human avatars, such as spiders or bats, and experience unfamiliar body structures as their own~\cite{krekhov2018vr}. In collaborative tasks, participants adapted to novel bodily forms and developed new communication strategies~\cite{espositi2025alien}. Experiencing natural environments through animal avatars has further been shown to increase empathy toward nature~\cite{ahn2016experiencing}, highlighting that embodiment transformation extends beyond visual substitution and influences cognition and values.  
Recent work has explored technological extensions of this flexibility. Neural networks have been used to translate human motion into quadrupedal forms such as dogs~\cite{egan2023neurodog, egan2024dog}, to embody non-human-like structures such as tentacles~\cite{takashita2024embodied}, and to leverage manual dexterity for controlling avatars of seahorses and elephants~\cite{jiang2023handavatar}. Kilteni et al. demonstrated that participants could control multi-armed avatars relatively naturally in VR, providing empirical support for Homuncular Flexibility. Collectively, these studies show that human embodiment is more adaptive and extensible than traditionally assumed.  
At the same time, research has revealed its boundaries. Zhang et al. investigated the psychological effects when professional dancers performed dance using avatars with different embodiments, such as avatars seated in a wheelchair~\cite{zhang2025becoming}, finding that reduced embodiment made it harder to link intentions to avatar control. This suggests inherent limits to bridging across different embodiments.  

Taken together, prior work highlights both the flexibility and the limits of embodiment. Building on this recognition, our work aims to extend embodiment theory beyond the notion of flexibility toward a framework that preserves and transfers intent across embodiments. To this end, we design BRIDGE, which integrates an embodiment-aware visualization framework with a pipeline for reconstructing sports video in three-dimensional space, and we empirically evaluate its effectiveness to demonstrate this new direction.

\subsection{Sports Visualization for Understanding Play}

The visualization of play in sports has evolved into a central approach for understanding the complex dynamics of team behavior~\cite{naik2022comprehensive, witte2025sports}. With the advancement of tracking technologies, a wide range of data-driven methods have been developed to represent player trajectories, passing flows, formations, and spatial coverage~\cite{gudmundsson2017spatio, torres2022tracking}.

In HCI, interactive visualizations have been actively explored to support both spectators and learners. Omnioculars overlays tactical indicators on game videos to enhance comprehension~\cite{lin2022quest}, while IBall leverages gaze-based interaction to improve immersion and understanding~\cite{zhu2023iball}. Sportify integrates action detection and tactical classification with embedded narratives to foster deeper engagement~\cite{lee2024sportify}. VisCourt applies Mixed Reality for in-situ tactical learning~\cite{cheng2024viscourt}, and Augmented Coach employs volumetric data for spatial annotation in coaching contexts~\cite{wen2024augmented}.

Meanwhile, foundational technologies in sports science and computer vision have advanced rapidly. GPS-, optical-, and IMU-based tracking systems are now widely used for analyzing external load and formations~\cite{torres2022tracking}. Advances in 3D mesh tracking have enabled not only skeleton-level but also surface-level body reconstruction, achieving practical accuracy for sports-specific pose estimation~\cite{fukushima2024potential}. Deep learning-based mesh recovery further supports multi-player and clothing-variant scenarios, with promising implications for analyzing complex interactions and contact plays~\cite{liu2024deep}.

Despite these advances, most visualization research has focused on non-disabled sports, leaving parasports such as wheelchair basketball largely unexplored. To address this gap, our work reinterprets plays from non-disabled athletes through embodied transformation and proposes a novel visualization framework that preserves tactical actions while adapting to the unique constraints of wheelchair sports. In doing so, this study contributes new perspectives to the design of inclusive learning environments.

\section{FORMATIVE STUDY}
\label{sec:formative}
We conducted a formative study to better understand the challenges players face when interpreting stand-up basketball footage in the context of wheelchair basketball. Through this study, we identified three recurring difficulties: (1) Difficulty in Interpreting Tactical Movements under Wheelchair-Specific Constraints, (2) Interpreting Body Orientation and Pose, and (3) Team-Level Tactical Alignment Across Functional Differences.

\begin{figure}[t]
    \centering
    \includegraphics[width=1.0\linewidth]{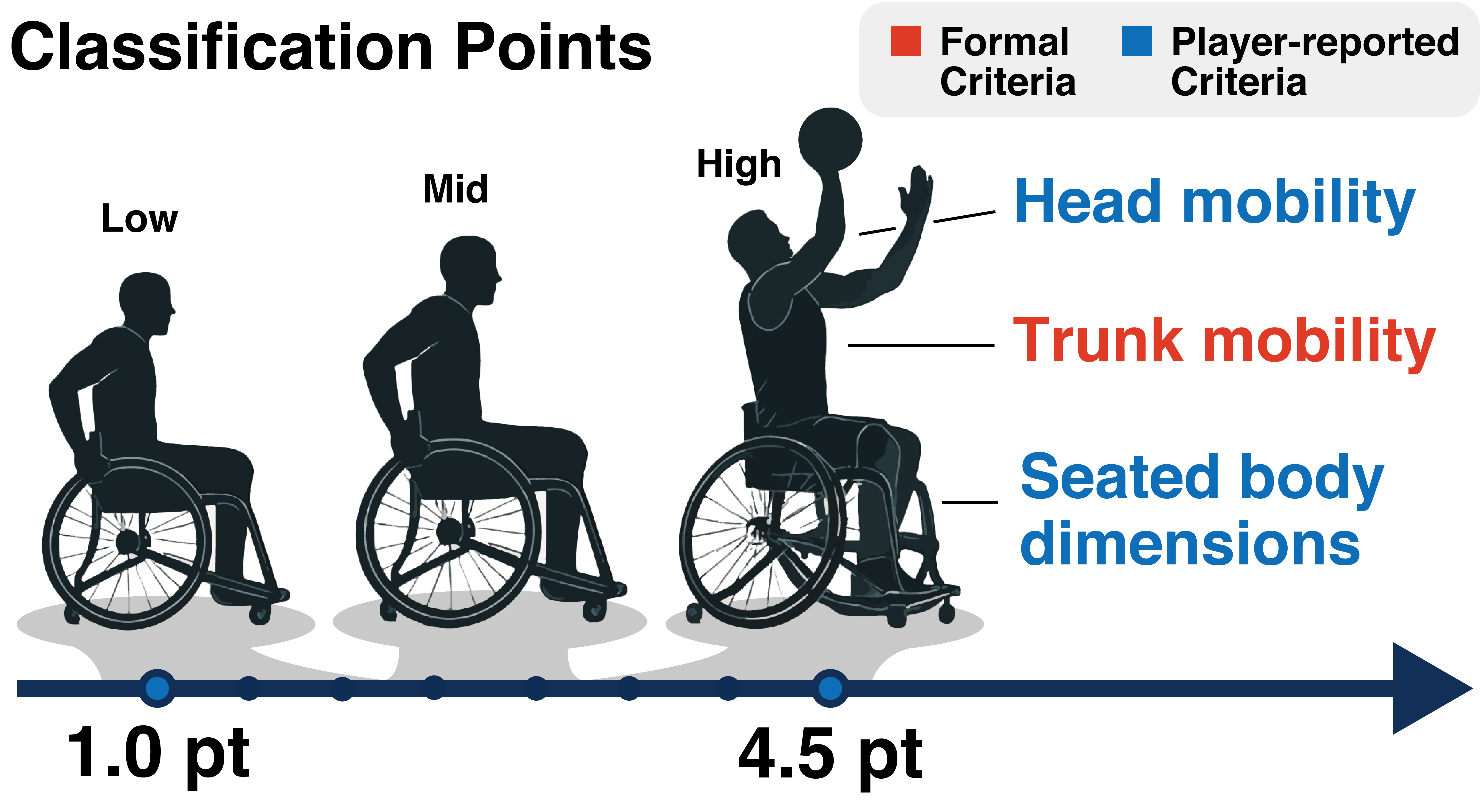}
    \caption{Classification points in wheelchair basketball. IWBF’s formal criteria (red) emphasize trunk mobility in the 1.0–4.5 point system, while players and coaches in our formative study additionally highlighted head mobility and seated body dimensions (blue) as decisive factors.}
    \Description{Illustration of wheelchair basketball classification points from 1.0 to 4.5. The figure shows wheelchair player silhouettes progressing from low to high trunk mobility along the point scale. Elements shown in red indicate the IWBF formal criteria emphasizing trunk mobility, while elements shown in blue represent additional factors reported by players and coaches, including head mobility and seated body dimensions.}
    \label{fig:Classification}
\end{figure}

\subsection{Functional Classification in Wheelchair Basketball} \label{classification}
Before presenting our formative study, we take into account the functional classification system defined by the International Wheelchair Basketball Federation (IWBF)~\cite{IWBF_classification_basics}.
As shown in Fig.~\ref{fig:Classification}, athletes are classified primarily based on trunk function, on a scale ranging from 1.0 to 4.5 in increments of 0.5. On the one hand, low-point athletes (e.g., 1.0–2.0) have little to no voluntary trunk control and therefore face greater physical constraints. On the other hand, high-point athletes (e.g., 3.5–4.5) retain near-complete trunk function and mobility.
According to IWBF rules, the total classification points of the five players on court must not exceed 14.

\subsection{Participants}

Based on the classification system described in Section~\ref{classification}, we recruited eight players from the national wheelchair basketball team and two national team coaches. All participants had extensive competitive experience at the national or international level and had previously attempted to incorporate insights from stand-up basketball videos into their own practice, ensuring direct relevance to this study. To represent a range of classification points, two athletes were included from each of the IWBF 1.0, 2.0, 3.0, and 4.0 classification points. The players had an average of 10.4 years of competitive experience (SD = 3.7), with four men and four women. The two coaches, one male and one female, both had prior coaching experience with the national team.

\subsection{Procedure} 
Each session was conducted individually online and lasted approximately 45 minutes. The first 15 minutes were devoted to a semi-structured interview, where participants were asked about their daily training methods, use of video materials, and challenges when learning from videos of non-disabled players, following a pre-prepared interview guide. The interviews were recorded using a PC screen recording tool and a smartphone audio recording application, with supplementary notes taken as needed.  
Next, four short tactical video clips created from YouTube videos were presented, including one shooting action from a \textit{screen}\footnote{A blocking play in which an offensive player positions themselves to impede a defender, freeing a teammate for a shot or drive.} (specifically, a \textit{flare screen}\footnote{A type of screen where the screener angles away from the basket to free a shooter on the perimeter.}), two actions using \textit{pick-and-roll}\footnote{A play in which a screener sets a screen on the ball handler’s defender and then moves (rolls) toward the basket to receive a pass.}, and one shooting action from an \textit{elevator screen}\footnote{A play where two screeners close together like doors after a shooter runs between them.} (average length 7.03 seconds, SD = 0.92)~\cite{youtubeNBA2022, youtubePacersPickRoll2014, youtubeWarriorsElevator2013, youtubeWarriorsWeave2015}.  
After each clip, participants were asked to reflect on how the depicted actions could be interpreted and applied in the context of wheelchair basketball. They were also encouraged to consider how their own \textit{functional differences} might shape these interpretations. Follow-up questions probed which elements seemed feasible, which required modification, and which appeared unsuitable for wheelchair basketball. Finally, a 15-minute closing interview was conducted to gather overall impressions and suggestions on how videos of non-disabled players could be made more useful for wheelchair basketball athletes.  

All interviews were conducted with participant consent and transcribed for analysis. 
Two authors independently coded the transcripts using grounded theory~\cite{muller2012grounded, muller2014curiosity}, focusing on recurring themes and difficulties. 
The codes were iteratively refined through discussion until consensus was reached. Inter-coder agreement, measured using Cohen’s Kappa, was 0.89.

\subsection{Interview Findings}
\subsubsection{Critical Role of Video in Daily Wheelchair Basketball Training}
In the interviews, most participants described video as playing a crucial role in their daily training. Many players reported learning tactical development and positioning by watching footage of top athletes from both domestic and international competitions. One participant noted, \textit{``When I watch videos of strong players, I can see what I’m missing''} (P4).
The videos they used included both wheelchair basketball and stand-up basketball games. The main reasons for using stand-up basketball footage were the lack of training materials specific to parasports and the desire to gain new insights. Official match recordings and technical explanations for wheelchair basketball were limited, leading some to comment, \textit{``We have no choice but to watch non-disabled players’ games instead''} (P3). At the same time, videos were seen as a valuable source of new knowledge and strategies: \textit{``Watching overseas games helps me understand new tactical trends''} (P10), and \textit{``There are useful hints even in the complex movements of stand-up basketball''} (P7). For many athletes, video watching had become an established part of everyday learning, serving not merely as imitation but as a means of deepening understanding.
However, several participants noted difficulties interpreting the intentions or tactical decisions in stand-up basketball footage due to differences in mobility and reach. They expressed that \textit{``it’s impossible to make the same movements in a wheelchair''} and that trying them often \textit{``feels completely different from what was expected''} (P1, P4, P6). Consequently, players repeatedly engaged in a cycle of \textit{watching, trying, and reviewing}, spending significant time to internalize the techniques in ways that matched their own physical sense (P4, P9).
Moreover, limited player numbers and training environments further restricted opportunities to apply what they learned from video. Participants mentioned challenges such as \textit{``there’s no one to practice with''} and \textit{``few gyms accept wheelchair basketball''} (P5, P7). As a result, video viewing played an even more central role in their practice and learning.

Given these circumstances, supporting wheelchair basketball players in learning effectively from stand-up basketball footage requires new approaches that help them intuitively interpret and relate such play to the context of wheelchair basketball.

\subsubsection{Difficulties in Interpreting Non-disabled Sports Footage}\label{Findings}

Based on the interview results, we identified three major gaps that wheelchair basketball players face when attempting to interpret non-disabled sports footage.

\textbf{D1. Difficulty in Interpreting Tactical Movements under Wheelchair-Specific Constraints.} 
Participants reported that it was difficult to imagine whether the movements shown in stand-up basketball footage could realistically be reproduced in wheelchair basketball.
For example, P3 noted, \textit{``From the video alone, I can see what is being done, but it is difficult to understand how I should move in order to fully reproduce it in a wheelchair.''}
Many participants (P1, P3, P5, P6, P8) similarly pointed out that movements often seen in stand-up basketball—such as rapid accelerations, sharp directional changes, and crossings in narrow spaces—were difficult to envision or execute under wheelchair play conditions.
Furthermore, some participants (P5, P6) mentioned that this difficulty in imagination hindered their ability to focus on tactical details. P6 specifically stated, \textit{``considerable effort was consumed merely imagining whether the movements were realistically possible in a wheelchair, which prevented me from focusing on tactical details.''}

These observations suggest that translating the movements of non-disabled players into wheelchair-specific contexts imposes a substantial cognitive burden.
As a result, participants found it challenging both to conduct detailed tactical analyses and to clearly envision how such movements could be practically executed in wheelchair play.

\textbf{D2. Interpreting Body Orientation and Posture.}
The treatment of body orientation fundamentally differs between stand-up and wheelchair basketball.
In stand-up basketball, during side steps or back steps, a player’s body orientation often does not align with their direction of movement.
In contrast, in wheelchair basketball, body orientation is closely tied to the direction of wheelchair movement, requiring players to remain constantly aware of their wheelchair’s facing direction (P1, P4, P9, P10).
However, when watching stand-up basketball footage, participants reported difficulty interpreting such movements. As P4 explained, \textit{``even if a non-disabled player twists their body, it does not necessarily indicate a change in wheelchair direction, so sometimes I do not know how to interpret it.''}  
This difference in bodily structure makes it challenging to interpret postures and movement cues accurately.
Thus, while participants could understand the tactical intentions behind the actions, physical differences introduced uncertainty in how those movements should be interpreted or executed (P1, P9).
Prior research has shown that body and facial orientation significantly affect the understanding of play intentions and passing success rates\cite{triggs2025perceptual}.
Therefore, the inability to accurately read directional cues poses a major limitation for tactical understanding when learning from video footage.

\textbf{D3. Team-Level Tactical Alignment Across Functional Classification Points Differences.}  
While non-disabled sports videos are useful for understanding tactical actions, they fail to account for the executional differences that exist across functional classification points.
This limitation makes it difficult to integrate and apply tactics collectively at the team level.
In wheelchair basketball, stability and mobility vary greatly across classifications; therefore, even when adopting the same tactic, the way it is executed differs significantly depending on the player’s classification (P2, P6, P7, P10).
One participant (P6) noted, \textit{``Even when applying the same tactic, the method of execution differs entirely across classes, and it is hard to coordinate and align movements with others during team play.''}
Another participant (P10) pointed out a communication-related challenge: \textit{``When sharing the same tactic with players from different classes, I sometimes can’t tell if my mental image is being accurately conveyed.''}

In summary, while players could understand the tactical actions themselves, video materials alone could not adequately represent the executional diversity arising from functional differences.
This diversity made it difficult to achieve tactical synchronization among players from different classifications.

\subsection{Functional Differences across Classification Points}

The formative study revealed that all participants consistently identified trunk mobility as the primary factor distinguishing classification points, with seated height and head mobility also highlighted as important indicators (Fig.~\ref{fig:Classification}).
Participants explained that \textit{low-point players} generally have little to no voluntary trunk control, relying on their arms for stability. As a result, they tend to sit deeper in the wheelchair, which lowers their seated height and center of mass. Some participants also mentioned that low-point players may have limited head mobility.
Regarding \textit{mid-point players}, participants described that they have partial trunk rotation and limited forward flexion, yet still experience substantial physical constraints compared to higher classes.
Finally, \textit{high-point players} were described as having nearly full trunk mobility, with higher seated height and center of mass that allow for greater freedom of movement.

Overall, trunk and head mobility, together with seated body dimensions, emerged from participants’ descriptions as the primary indicators distinguishing functional capacities across classification points.
\begin{figure*}[t]
    \centering
    \includegraphics[width=1.0\linewidth]{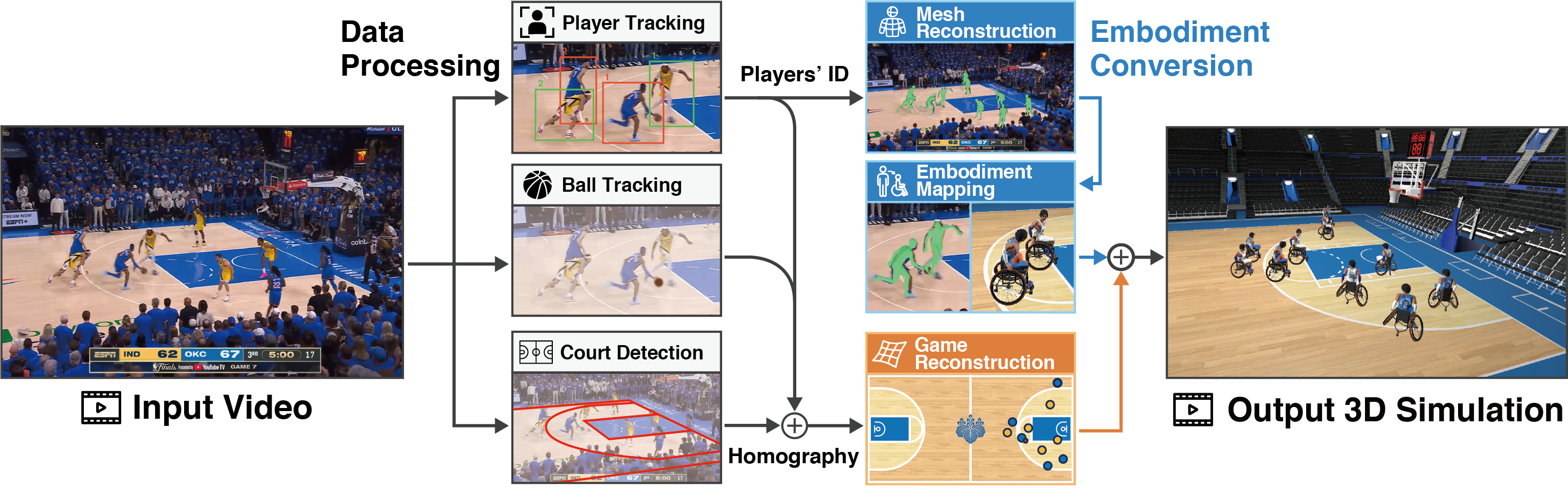}
    \caption{System pipeline overview: Input video is processed through player, ball, and court detection, followed by pose estimation and mesh reconstruction. Embodiment mapping and game reconstruction are then integrated to reconstruct the wheelchair basketball scene in a 3D simulation space. Images include still frames from NBA broadcast footage used for research illustration purposes. © NBA / Broadcast rights holders.}
    \Description{Overview of the BRIDGE system pipeline. The diagram shows how broadcast basketball video is processed through player, ball, and court detection, followed by pose estimation and mesh reconstruction. These outputs are then combined through embodiment mapping and game reconstruction to generate a 3D wheelchair basketball simulation.}
    \label{fig:system}
\end{figure*}

\section{SYSTEM DESIGN}
\label{sec:design}

\subsection{Design Goals}
\label{sec:Goals}
Based on the findings from our formative study, we established the following design goals to guide the development of our system. 

\textbf{G1. Facilitate the imagination of wheelchair execution:} Our system enables users to more easily imagine how movements depicted in stand-up basketball videos could be executed in wheelchair basketball. By supporting the translation of actions from non-disabled players into wheelchair-specific contexts, the system can reduce the effort spent on feasibility judgments.
This enables a more actionable understanding of tactical details such as wheelchair orientation, timing of feints, and coordination with teammates.

\textbf{G2. Support an accurate interpretation of body orientation and posture:} Our approach provides a mapping of the head, trunk, and wheelchair base tailored to the mobility needs of wheelchair basketball.
This reduces the ambiguity in stand-up basketball videos where body orientation does not correspond to movement direction. This mapping helps players better understand the relationship between body orientation and wheelchair movement, allowing them to interpret tactical actions more accurately from posture and orientation cues. 

\textbf{G3. Enable classification-aware team-level tactical adaptation:}
This design highlights how tactical execution differs across functional classes, addressing the limitation that video materials alone cannot capture these variations. By reflecting differences in stability, mobility, and range of motion, it supports the coordination and alignment of strategies across players with different classifications, enabling more effective team-level tactical adaptation.

\subsection{System Overview} 
Our system reconstructs stand-up basketball footage into a 3D wheelchair basketball simulation through a multi-stage pipeline, as illustrated in Fig.~\ref{fig:system}. The pipeline consists of two main modules. First, the \textit{Game Streaming Reconstruction} module processes broadcast stand-up basketball footage, detecting players, the ball, and court geometry, and mapping trajectories into a virtual 3D court environment. Second, the \textit{Embodiment-aware Orientation Mapping} module reinterprets body orientation cues by separating the upper body (including the head) from the wheelchair base. This mapping enables the action affordances, tactical actions, and feasible movement directions in wheelchair basketball. Together, these components address the challenges identified in our formative study and realize the design goals outlined in Section~\ref{sec:Goals}.

\subsection{Game Streaming Reconstruction (Supports G1)}
\label{sec:Reconstruction}
The first stage of our pipeline reconstructs broadcast stand-up basketball footage into a virtual 3D court environment within Unity. This process involves several key steps: \textbf{1) Player and Ball Detection and Tracking}, \textbf{2) Court Keypoint Detection, Homography, and Position Mapping}, \textbf{3) Ball Possession and Action State Detection}, and \textbf{4) Trajectory Smoothing}. Together, these steps produce realistic reconstructions of gameplay, which are then instantiated as wheelchair player models to support orientation mapping (Section~\ref{sec:Mapping}) and enable tactical scenarios from stand-up basketball footage to be faithfully represented in a virtual wheelchair basketball environment. Ultimately, this 3D game reconstruction pipeline enables users to envision how movements depicted in stand-up basketball footage could be executed in wheelchair basketball, thereby supporting G1.

\textbf{1) Player and Ball Detection and Tracking:}  
To reliably identify players and the ball across frames, we adopted detection and tracking methods. For players, we employed \textit{MixSort}~\cite{cui2023sportsmot}, a tracking-by-detection framework that combines motion-based association (via Kalman filtering and IoU matching) with appearance-based association through a transformer-like module. This hybrid design is particularly advantageous in dynamic scenes such as basketball plays, where players frequently overlap, share similar uniforms, and move at variable speeds. On the SportsMOT benchmark, MixSort achieved 65.7 HOTA and 74.1 IDF1 overall, with basketball being the most challenging category (60.8 HOTA, 67.8 IDF1).\footnote{HOTA (Higher Order Tracking Accuracy) and MOTA (Multiple Object Tracking Accuracy) are standard benchmarks commonly used in multi-object tracking. IDF1 is an F1-score variant reflecting identity preservation in tracking.} By adopting MixSort, we obtained consistent player IDs across frames, providing temporally stable bounding boxes for subsequent spatial reasoning and tactical analysis. These stable tracking IDs also allowed us to map players into the reconstructed 3D space, ensuring continuity across sequences. For the ball, whose small size, rapid movement, and frequent occlusions posed unique challenges, MixSort was not suitable. We employed \textit{YOLOv10}~\cite{wang2024yolov10} for detection to provide reliable prompts and integrated \textit{SAM2}~\cite{ravi2024sam} for multi-frame association. This combination ensured accurate, frame-by-frame localization of the ball, which was critical for analyzing possessions and player–ball interactions.

\textbf{2) Court Keypoint Detection, Homography, and Position Mapping:}  
To enable consistent spatial reasoning across broadcast videos, we first incorporated a YOLO-based detector to recognize key structural landmarks of the basketball court, such as sidelines, baselines, free-throw lines, and three-point arcs. These landmarks served as geometric anchors for computing a homography transformation from the broadcast image plane to a canonical court template. Establishing this homography mitigated perspective distortion and provided a normalized spatial model. With this reference frame in place, player and ball bounding boxes could then be projected into standardized court coordinates, converting raw image-space detections into interpretable positions. This step was essential for analyzing movements and tactical patterns in a coherent space, as well as synchronizing trajectories across sequences despite varying camera viewpoints, thereby ensuring that all detected entities were consistently located on the court.

\textbf{3) Ball Possession and Action State Detection:}  
Based on the previously described player and ball tracking, we developed a mechanism to detect ball possession. Specifically, when more than 70\% of the ball’s bounding box (bbox) overlapped with the bounding box of an offensive player, that player was judged to be in possession of the ball. To account for detection errors, this condition had to be satisfied for at least five consecutive frames before possession was confirmed.
When ball possession shifted from one player to another, the event was classified as a “pass,” and the interval during which the ball was in transit was recorded as the pass trajectory. Furthermore, when the ball remained separated from all players at the end of the data sequence, the event was regarded as a “shot.”
Based on these detection results, the system automatically classified the ball possessor’s actions into “pass,” “dribble,” or “shot,” and assigned corresponding animations. This process enabled a more faithful reconstruction of in-game dynamics and supported users in assessing the feasibility and tactical implications of the reconstructed plays.
However, not all movements observed in stand-up basketball can be directly reproduced in wheelchair basketball (e.g., jumping, stepping, or single-leg balance actions).
In such cases, our system approximates these movements as feasible wheelchair actions based on the acquired positional data, movement direction, and action state of each player.
Furthermore, different types of shots and passes (e.g., hook shots\footnote{A one-handed shot in which the player extends the shooting arm sideways and releases the ball in a sweeping motion over the shoulder, allowing it to arc over a defender.}, floaters\footnote{A high-arching shot, usually taken by a smaller player driving toward the basket, designed to go over taller defenders.}, one-handed passes, and bounce passes) are abstracted and represented as representative wheelchair basketball actions—such as standard shots or chest passes—while preserving their tactical meaning.
This approach ensures that variations in biomechanical characteristics do not affect the interpretability or consistency of the overall tactical intent and flow of play.

\textbf{4) Trajectory Smoothing:}  
Finally, to address residual noise, jitter, and short-term occlusion errors in the tracking outputs, we applied a Kalman filter–based smoothing procedure. This step produced more continuous and realistic motion trajectories, which were essential for faithfully reconstructing the dynamics of gameplay. The reconstructed trajectories were then instantiated as wheelchair player models within Unity, forming the basis for subsequent orientation mapping (Section~\ref{sec:Mapping}). This reconstruction ensures that tactical scenarios from stand-up basketball footage can be faithfully represented in a virtual wheelchair basketball environment.

\begin{figure*}[t]
    \centering
    \includegraphics[width=1.0\linewidth]{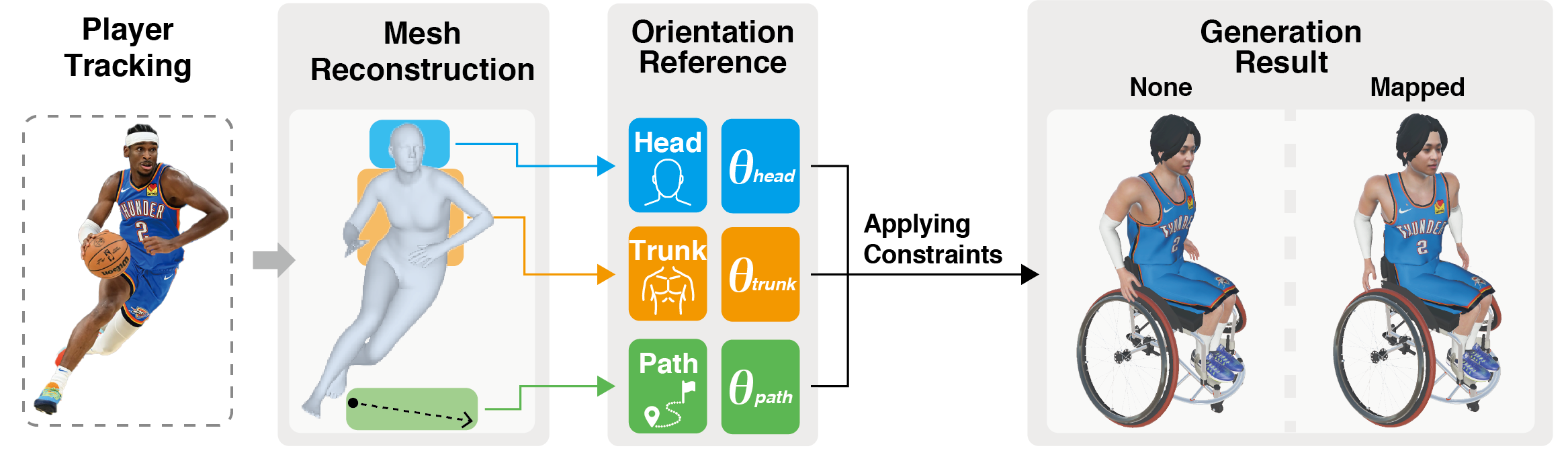}
    \caption{Overview of the embodiment-aware orientation mapping process. Player tracking provides mesh reconstruction, from which head, trunk, and path orientations are referenced. By applying classification-based restrictions, the system generates wheelchair basketball player models. The right panel compares results without mapping and with mapping. Images include a player photograph used for research illustration purposes. © Joshua Gateley / Getty Images.}
    \Description{Diagram illustrating the embodiment-aware orientation mapping process. The figure shows how player tracking data is used to reconstruct a mesh and extract head, trunk, and movement path orientations, which are then constrained by classification-based rules to generate wheelchair basketball player models. The right side compares player representations without orientation mapping and with orientation mapping applied.}
    \label{fig:Direction}
\end{figure*}

\subsection{Embodiment-aware Orientation Mapping (Supports G2, G3)}
\label{sec:Mapping}
In this section, we present an embodiment-aware orientation mapping model that translates movements from stand-up basketball to wheelchair basketball by incorporating classification-based constraints (Fig.~\ref{fig:Direction}). In this model, the wheelchair base defines the primary heading, the trunk is constrained around the base with a classification-dependent range of motion, and the head is further constrained relative to the trunk. This hierarchical design allows us to represent trunk and head mobility more faithfully, while ensuring tactical cues remain interpretable within the embodiment of wheelchair basketball. Moreover, by encoding classification-dependent ranges of motion, the model directly supports both G2 and G3.

\textbf{1) Functional Motivation and Morphological Constraints:}

The movements observed in stand-up basketball players cannot be directly applied to wheelchair basketball. Therefore, in this study, we propose a framework that extracts the direction of motion, trunk orientation, and head orientation from stand-up basketball movements, and represents them within a three-layer hierarchical structure consisting of the \textit{wheelchair base}, \textit{trunk}, and \textit{head}. Among these layers, the trunk and head were emphasized by athletes during our formative study as critical factors. In particular, it was revealed that low-point players experience significant limitations in trunk control and postural stability. Consequently, by constraining trunk rotation relative to the wheelchair base, it becomes possible to model both the physical limitations and their tactical implications with greater precision.

\textbf{2) Setting Functional Differences Across Classifications:}  
In order to parameterize the hierarchical model according to player classifications, we first examine empirical evidence on functional differences. Prior studies have reported that trunk rotation among wheelchair basketball athletes classified from 1.0 to 4.5 points averages approximately $33^\circ$, with a range of about $20^\circ$ to $45^\circ$ across participants~\cite{marszalek2019reliability}. The lower bound of $20^\circ$ corresponds primarily to low-point players and reflects maximal effort under controlled testing conditions. During actual gameplay, athletes with severely limited trunk stability cannot fully utilize this range, as doing so increases the risk of collapse or loss of balance. Therefore, in this study, we pragmatically set the maximum trunk rotation for low-point players to approximately $10^\circ$. In contrast, high-point players exhibit trunk stability comparable to non-disabled individuals and can safely achieve up to about $45^\circ$, which we adopt as their upper bound.
Head orientation is modeled relative to the trunk, and its flexibility also depends on player classification. For high-point players, we refer to prior research indicating that healthy individuals can achieve up to approximately $80^\circ$ of head rotation~\cite{ferrario2002active}. In contrast, spinal cord injuries, which are common among wheelchair users, are known to affect head mobility, yet few studies have systematically examined classification-dependent differences. Based on reports gathered during our formative study, we set the maximum head rotation for low-point players to $45^\circ$. This value corresponds to about 60\% of the range observed in high-point players, thereby reflecting classification-dependent functional mobility differences while preserving the tactical importance of head orientation.

\textbf{3) Mathematical Formulation:}
To implement the proposed hierarchical mapping in a systematic way, we provide a mathematical formulation that specifies how orientations are transformed.

\[
\begin{cases}
\theta_{base} = f_{motion}(\text{traj}) \\
\theta_{trunk} = \theta_{base} + \text{clip}\bigl(\alpha(p)(\theta_{trunk}^{raw}-\theta_{base}), -\Delta^{\max}_{trunk}(p), \Delta^{\max}_{trunk}(p)\bigr) \\
\theta_{head} = \theta_{trunk} + \text{clip}\bigl(\beta(p)(\theta_{head}^{raw}-\theta_{trunk}), -\Delta^{\max}_{head}(p), \Delta^{\max}_{head}(p)\bigr)
\end{cases}
\]

Here, $\theta_{base}$ is computed from the motion trajectory and aligned with the instantaneous displacement vector $(x_{t} - x_{t-1}, y_{t} - y_{t-1})$. 
$\theta_{trunk}^{raw}$ and $\theta_{head}^{raw}$ denote the original torso and head orientations extracted from the stand-up basketball footage, serving as unconstrained reference angles. $\theta_{trunk}$ is then defined relative to this base orientation but constrained within a classification-dependent maximum range $\Delta^{\max}_{trunk}(p)$, reflecting the physical limits of torso rotation relative to the wheelchair base. Similarly, $\theta_{head}$ is defined relative to the trunk and constrained within $\Delta^{\max}_{head}(p)$, which represents the feasible range of head mobility. The coefficients $\alpha(p)$ and $\beta(p)$ modulate the extent to which the trunk and head orientations follow the raw input relative to their parent segments, thereby encoding classification-specific mobility constraints.

\textbf{4) Orientation Estimation:}  
To apply the hierarchical orientation model defined in the previous section to real player data, it is necessary to first estimate the posture of each athlete. For this purpose, we leverage the player tracking and identification described in Section~\ref{sec:Reconstruction}, and extract SMPL mesh information for each player using the \textit{mLCoMotion} framework~\cite{newell2025comotion, loper2023smpl}. 
From these mesh representations, we derived hierarchical orientation parameters (base, trunk, and head) and applied them to our wheelchair basketball models. This allowed us to ground orientation cues in robust pose estimation while ensuring consistency with embodiment constraints. According to the original evaluation, mL-CoMotion achieved over 70\% MOTA on PoseTrack21 and around 60\,mm MPJPE on 3DPW, demonstrating state-of-the-art accuracy at the time of publication.

Furthermore, by aligning the wheelchair base and trunk with the direction of motion and constraining the head relative to the trunk, our model can reinterpret movements such as side steps or back steps—actions infeasible in wheelchair basketball—as smooth heading adjustments. This enables continuous and realistic forward motion without abrupt changes. In addition, by representing situations in which players look left or right while moving forward in a manner consistent with the original video, the system allows more accurate inference of gaze direction and tactical action, thereby preserving the meaning of strategic decisions during gameplay.

\subsection{Processing Time and Implementation}
To quantitatively evaluate the computational efficiency of our pipeline, we analyzed five basketball play videos (average length 7.6 ± 1.5 seconds, 228 ± 44.9 frames). The processing pipeline consists of three main modules: player tracking using YOLOX + MixSort, ball tracking with SAM2, and tactical view conversion with drawing. The average total processing time per video was 85.4 ± 22.8 seconds. The module-wise processing speeds were 11.5 ± 0.5 fps for player tracking, 13.2 ± 0.8 fps for 3D pose estimation, and 3.6 ± 0.5 fps for ball tracking. Tactical view conversion was computationally inexpensive (669 ± 102 fps). These results provide a comprehensive overview of the system’s computational characteristics under realistic conditions.
Experiments were conducted on a workstation running Windows 11 with an Intel Core i7-13700K CPU, an NVIDIA GeForce RTX 4090 GPU (16 GB VRAM, driver version 571.96, CUDA 12.8), and 32 GB RAM.

Furthermore, to visualize and interactively reproduce the processed results, we employed Unity 6000.0.23f1 as the integration environment. Player position information obtained through homography transformation and wheelchair player posture information estimated via embodiment mapping were synchronously reconstructed in Unity, enabling the dynamics of plays to be reproduced in three-dimensional space. In addition, animations specific to wheelchair basketball, such as pushing, passing, and shooting, were generated using motion capture data recorded from members of the Japanese national team, thereby achieving a more realistic reproduction of actual gameplay.

\section{USER STUDY}
It is essential that BRIDGE preserves the same tactical actions as stand-up basketball footage and functions as a support to facilitate understanding of how such actions can be executed under embodiment constraints and wheelchair-specific limitations. To evaluate the effectiveness of BRIDGE, we conducted two user studies with wheelchair basketball players. In the first study, we examined how the proposed embodiment-aware orientation mapping influences perceptions of motion realism and functional differences. In the second study, we compared broadcast footage of stand-up basketball with reconstructed wheelchair basketball footage to investigate whether tactical action is preserved across embodiments and how it affects subjective evaluation items such as self-efficacy and ease of imagery. Finally, we conducted interviews to collect qualitative insights, including participants’ perceptions, subjective feedback, and suggestions for improvement.

\begin{table*}[h]
\centering
\caption{Participants’ demographics}
\Description{Table summarizing participant demographics for national team and non-elite wheelchair basketball players, including gender, age, playing experience, disability status, and classification scores.}
\label{tab:participants}
\begin{tabular}{lcccccc}
\toprule
Group & N & Gender & Age & Wheelchair BB Experience & Disability & Classification \\
      &   & (M/F)  & (M$\pm$SD) & (M$\pm$SD) & (Yes/No) & (M$\pm$SD) \\
\midrule
National team & 10 & 6/4  & 31.6$\pm$4.01 & 12.9$\pm$6.08 & 10/0 & 3.0$\pm$1.2 \\
Non-elite       & 10 & 4/6  & 20.8$\pm$2.25 & 1.5$\pm$0.88  & 2/8  & 2.75 \\
Overall       & 20 & 10/10 & 26.2$\pm$6.38 & 7.2$\pm$7.22 & 12/8 & 2.96$\pm$1.14 \\
\bottomrule
\end{tabular}
\end{table*}

\subsection{Participants and Video Preparation}
This study involved 20 participants, including 10 members of the Japanese national wheelchair basketball team and 10 non-elite players, as shown in Table~\ref{tab:participants}.
All participants had prior experience in wheelchair basketball, and three also had experience playing stand-up basketball. Novices were excluded to avoid confounds arising from a lack of basic knowledge of the sport. The non-elite group consisted mainly of players affiliated with university or community clubs, eight of whom were non-disabled. These participants regularly played wheelchair basketball together with players with disabilities on the same teams and routinely used sports wheelchairs in both practice and competition. Due to the limited number of players with disabilities available for recruitment, we included non-disabled players who regularly train and compete in wheelchair basketball under official rules and wheelchair-use conditions.  All non-disabled participants were also familiar with the classification system and rules of wheelchair basketball.
By including both national and non-elite players, the sample represented a broad range of competitive expertise. The study was approved by the institutional ethics review board. All participants provided written informed consent and received compensation upon completion.

Video stimuli were created from YouTube footage of NBA games. Ten clips, each containing a single tactical play, were first extracted. After consultation with expert coaches, eight clips were selected ($M = 7.85$ seconds, $SD = 1.66$) to ensure both tactical validity and diversity~\cite{youtubeNBA2022, youtubePacersPickRoll2014, youtubeWarriorsElevator2013, youtubeWarriorsWeave2015, youtubeHawksThruAction2013, youtubeMavsPostUpDecoy2014, youtubeSpursHalfCourtSets2017, youtubeWarriorsHornsFlex2013}. A national-level coach classified the clips into four simple and four complex tactics. In Study 1, two simple and two complex tactics were used; the remaining clips were used in Study 2, ensuring balanced task complexity across studies.

All videos presented as wheelchair basketball footage were reconstructed as a three-dimensional space in Unity, and the same scenes were recorded within that space to generate the final videos.
To reflect functional differences identified in the formative study, the reconstructed videos depicted three player types. \textbf{Low-pointers} were shown in smaller wheelchairs with strong restrictions on trunk and head rotation, \textbf{mid-pointers} in medium-sized wheelchairs with moderate restrictions, and \textbf{high-pointers} without restrictions, allowing a range of motion equivalent to non-disabled players. These player types were assigned to each team in a 2:1:2 ratio (low:mid:high) and randomly distributed within the lineup.

\subsection{Study 1: Effects of Body Orientation Mapping on Users’ Understanding}

The purpose of this study was to examine how the proposed embodiment-aware orientation mapping influences the perception of postural naturalness and functional differences in the reconstructed footage.

\subsubsection{Procedure.} 
In the first part, participants viewed four reconstructed wheelchair basketball videos presented in randomized order to mitigate order effects and reduce potential bias. Each video was presented side by side in two conditions:
(mapped condition) the proposed embodiment-aware orientation mapping; and
(baseline condition) a condition that mapped only the movement direction to the wheelchair’s facing direction while leaving trunk and head orientations unmapped. Participants selected which version appeared more natural (“mapped,” “baseline,” or “no difference”) and then rated the naturalness of player postures in each condition using a 7-point Likert scale.

In the second part, participants again viewed the same videos, but only under the mapping condition. They estimated the functional classification of the five offensive players according to wheelchair basketball standards. They then rated the extent to which three factors identified in our formative study (body dimensions, trunk range of motion, and head range of motion) influenced their judgments on a 7-point Likert scale, and finally selected the single factor they considered most influential.

\subsection{Study 2: Comparing Stand-up Basketball Footage and Wheelchair Basketball Reconstructions}

The purpose of this experiment was to examine whether the tactical action is preserved when participants view reconstructed wheelchair basketball animations compared to original stand-up basketball footage. In addition, it was intended to investigate how this viewing experience affects participants’ self-efficacy and cognitive load during the interpretation of tactical plays.

\subsubsection{Procedure.}  
This study adopted a within-subject design in which each participant viewed four tactical plays under two conditions:
(S condition) original stand-up basketball footage; and 
(W condition) reconstructed wheelchair basketball footage.
Each condition included one simple and one complex tactic. The assignment of tactics was randomized for each participant to counterbalance order and content effects.
After viewing each video, participants were asked to evaluate the retention of tactical action. Following the procedure of Lee et al.,\cite{lee2024sportify} the task required participants to reproduce the sequence of actions by arranging cards labeled “screen,” “pass,” “cut,” and “shoot.” For each video, participants were instructed to reproduce the actions of a designated player.
During the task, we recorded both the number of video replays and the time participants spent reasoning before submitting their answers. The accuracy of their responses was assessed by comparing them with model answers prepared by a national-level coach. After completing the recall task, participants rated their self-efficacy and several subjective evaluation items related to the plays presented in the videos, using seven-point Likert scales.

After completion of all trials, the participants participated in a qualitative interview that assessed their overall impressions, perceived effectiveness of the system as a learning aid, and suggestions for improvement.

\subsubsection{Measures}
\textbf{Self-efficacy.}
We collected self-efficacy ratings to understand participants' confidence in their learning capacity
 following Bandura’s guidelines~\cite{ba06guide}. While self-efficacy is influenced by the four sources identified by Bandura (mastery experiences, vicarious experiences, social persuasion, and physiological feedback), vicarious experience is particularly relevant in the present study because it is closely tied to observational learning~\cite{bandura1982self}. In disability sports, the movements presented as learning models may differ from the observer’s own bodily form, and such discrepancies can affect the process of mapping observed actions onto one’s own sensorimotor context. Therefore, assessing self-efficacy provides an important indicator of how well participants can integrate and internalize knowledge and movements that have been transformed from a different modality into their own bodily modality.

\textbf{Subjective Evaluation Criteria.}
In addition, we collected multiple subjective evaluations on the overall learning experience:
Q1 \textit{imagery vividness}, Q2 \textit{ease of imagery}, Q3 \textit{usefulness for tactical learning}, Q4 \textit{usefulness for exploring new tactics}, and Q5 \textit{intention to use}. For the imagery items, participants were asked to imagine performing the plays themselves in a wheelchair.  
The meaning of each scale point was defined as follows: for Q1, 1 = no image at all, 4 = moderately clear and vivid, and 7 = perfectly clear and vivid; for Q2, 1 = very hard to imagine, 4 = neutral, and 7 = very easy to imagine; and for Q3–Q5, 1 = strongly disagree, 4 = neutral, and 7 = strongly agree.  
These items were adapted from existing questionnaires (e.g., VMIQ, SIAQ, TAM, UTAUT)~\cite{williams2011measuring, roberts2008movement, davis1985technology, venkatesh2003user}, with reference to prior research on sports training and exercise, and were modified to fit the context of wheelchair basketball learning~\cite{tsai2020feasibility, rosendahl2024360}.

   
\section{RESULTS}
In this section, we report the results of two user studies that examined how embodied visualizations and reconstructed wheelchair basketball videos influence perception, understanding, and self-efficacy. The findings are organized into five themes: (1) postural naturalness, (2) accuracy of functional classification and influencing factors, (3) preservation of tactical action, (4) self-efficacy, and (5) subjective evaluations. In addition, each section includes participants’ comments to highlight the implications of embodiment transformation for learning.

\begin{figure}[h]
    \centering
    \includegraphics[width=1.0\linewidth]{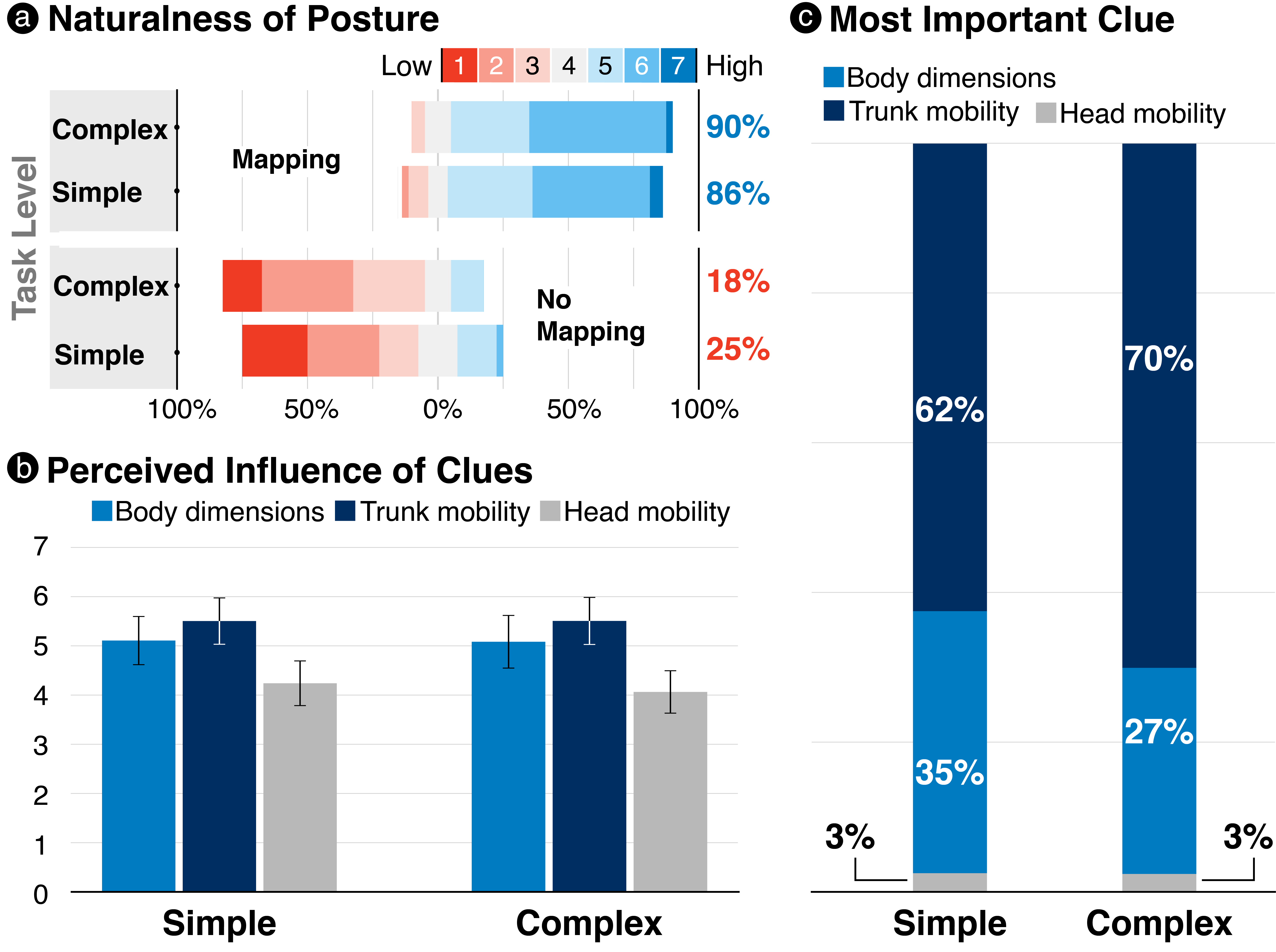}
    \caption{The results of User Study 1. (a) Participants rated player movements as more natural in the mapping condition than in the non-mapping condition when watching both Simple and Complex tactical videos. (b) When estimating player points, participants reported that trunk mobility had the greatest influence, followed by body dimensions and head mobility. (c) Participants identified trunk mobility as the most important clue for point estimation.}
    \Description{Results of User Study 1 shown in three subfigures. Subfigure (a) shows higher perceived naturalness ratings for player movements in the mapping condition compared to the non-mapping condition for both simple and complex tactical videos. Subfigure (b) indicates that trunk mobility was rated as the most influential factor for estimating player classification points, followed by body dimensions and head mobility. Subfigure (c) shows that participants most frequently identified trunk mobility as the primary cue for point estimation.}
    \label{fig:Study1}
\end{figure}

\subsection{Embodiment-aware Mapping Improves Perceived Naturalness}

We examined the results of Study 1 on how the proposed embodiment-aware orientation mapping influences the perceived naturalness of postures in reconstructed footage. 

When comparing the mapped and baseline conditions, the majority of participants overwhelmingly judged \textbf{the mapped condition as more natural} (mapped = 66, baseline = 3, no difference = 11). As shown in Fig.~\ref{fig:Study1}, Likert ratings on a 7-point scale confirmed the same trend: while few participants perceived the baseline condition as natural, nearly 90\% rated the mapped condition as natural. The mean ratings also differed, with the baseline condition at $M \approx 3$ and the mapped condition at $M \approx 5$.
A three-way ANOVA further confirmed that \textbf{the effect of mapping was statistically significant} ($p < .001$).
The effect size was large (partial $\eta^2=0.776$), 
and the post-hoc power analysis showed sufficient statistical power ($1-\beta=1.00$).

Participants’ comments reinforced these quantitative results. For the baseline condition, some described it as \textit{``feeling like a strategy board''} (P12, Non-elite) or noted a \textit{``robotic impression''} (P7, National). In contrast, for the mapped condition, participants emphasized that it \textit{``looked closer to an actual game''} (P3, National) and that \textit{``it was easier to grasp the intention of the play''} (P9, National). These observations suggest that \textbf{embodiment-aware mapping not only enhances perceived naturalness but also contributes to tactical understanding}.  

Nevertheless, not all participants perceived a difference. P11 (Non-elite) remarked that \textit{``even with fixed postures, it did not look unnatural,''} and P1 (National) pointed out that \textit{``for simple tactical understanding, wheelchair orientation alone can sometimes be sufficient,''} indicating that in certain contexts, the baseline condition can still be useful.

\subsection{Accurate Functional Classification through Controlled Cues}
We analyzed the classification accuracy of participants in reconstructed videos in Study 1 to examine how accurately and effectively orientation mapping, considering embodiment, can represent functional differences among players. 

\textbf{Participants classified offensive players into three categories (low-, mid-, high-pointers) with high accuracy}. Under the mapping condition, the average accuracy was $M = .80$ (95\% CI [.71, .89]) for simple plays and $M = .74$ (95\% CI [.65, .83]) for complex plays. A three-way ANOVA revealed no significant differences between player groups (National vs. Non-elite) or by complexity, and overall classification accuracy remained stable. 
However, the effect size was small, suggesting that the limited sample size may have led to low statistical power (partial $\eta^2 = .03$, $1-\beta = .18$). 
As shown in Fig.~\ref{fig:Study1}b and Fig.~\ref{fig:Study1}c, \textbf{the most influential cue for classification was consistently trunk mobility}. The three-way ANOVA revealed a significant main effect of Clue ($p = .004$, partial $\eta^2 = .26$, $1-\beta = .96$), indicating a large effect. On a 7-point scale, trunk ($M = 5.60$), body dimensions ($M = 5.37$), and head ($M = 4.30$) were reported for simple plays, and trunk ($M = 5.63$), body dimensions ($M = 5.30$), and head ($M = 4.17$) showed similar patterns for complex plays. Furthermore, trunk mobility was selected as “the most influential” cue in 62\% of cases for simple plays and 70.0\% for complex plays, with chi-square tests confirming significant differences across both conditions ($p = .0015$). These results align with participants’ comments such as: \textit{“Trunk rotation clearly represented each category and was most convincing”} (P3, National), and \textit{“The difference between players moving only the head and those also moving the trunk was clear and easy to judge”} (P7, Non-elite).  
Body dimensions also proved to be an effective supplementary cue, with several participants reporting that it was particularly useful for distinguishing low-pointers (P2, National; P13, Non-elite). However, some participants noted possible misjudgments: \textit{“When players came closer, they looked bigger, so I relied more on trunk mobility”} (P9, Non-elite).  
On the other hand, many found it difficult to distinguish between mid and high players. Comments included: \textit{“Even in actual games, they are often confused”} (P8, National), and \textit{“It was difficult since I don’t usually use point classifications in daily activities”} (P10, Non-elite). As for improvements, some participants suggested that point values should be made more explicit, for example through color coding or numbering (P5, P8, National), indicating the usefulness of strengthening visual support. 

Overall, participants were able to achieve stable and relatively high classification accuracy by primarily relying on trunk mobility, suggesting that \textbf{the mapping effectively conveyed functional differences in a visually comprehensible manner}.

\subsection{Tactical Action Preserved in Reconstructed Footage}

In Study 2, participants viewed both stand-up basketball footage (S condition) and reconstructed wheelchair basketball footage (W condition) to examine how accurately the reconstructed videos conveyed tactical action in wheelchair basketball.  

First, \textbf{regarding accuracy of tactical understanding, no major differences were observed} between the S and W conditions. For complex plays, accuracy was $M = .80$ in the S condition and $M = .90$ in the W condition. For simple plays, accuracy was $M = 1.00$ in the S condition and $M = .70$ in the W condition. Although slight differences were observed, the ART-ANOVA revealed no significant effects ($p > .05$, partial $\eta^2 = .06$, $1 - \beta = .62$), indicating that \textbf{tactical action was understood with high accuracy in both conditions}. In other words, reconstructed footage was sufficient to convey tactical action.  

Next, \textbf{the time required for understanding showed no significant differences} between conditions, with only task complexity having a significant main effect ($p = .031$, partial $\eta^2 = .08$, $1 - \beta = .83$). For complex plays, mean thinking time was 36.0 s in the S condition and 36.8 s in the W condition. For simple plays, it was 29.7 s in the S condition and 28.0 s in the W condition, indicating that more time was spent on complex tasks. Viewing frequency differed by player level ($p = .031$, partial $\eta^2 = .21$, $1 - \beta = 1.00$), with non-elite players watching on average 3.0 times compared to 2.7 times for national players, showing that less experienced participants needed to rewatch more frequently.

Participants’ comments further supported these findings. P8 (National) noted that \textit{``the overall flow and action of the plays were clearly understood in both videos.''} Similarly, P18 (Non-elite) remarked, “Since I usually do not watch stand-up basketball, seeing it reconstructed as wheelchair play made it easier for me to understand.” This suggests that the tactical action of the plays can be sufficiently understood in both the stand-up basketball footage and the reconstructed footage.
At the same time, some pointed out subtle differences. P4 (National) commented that \textit{``in the reconstructed videos, the execution of screens looked somewhat different from typical wheelchair basketball—still recognizable as screens, but slightly unusual,''} indicating that while tactical action was conveyed, there remained minor discrepancies in wheelchair angles and movement.  

In summary, \textbf{reconstructed footage was found to preserve tactical action as effectively as original footage,} with additional potential benefits for viewers less familiar with stand-up basketball.

\begin{figure*}[t]
    \centering
    \includegraphics[width=1.0\linewidth]{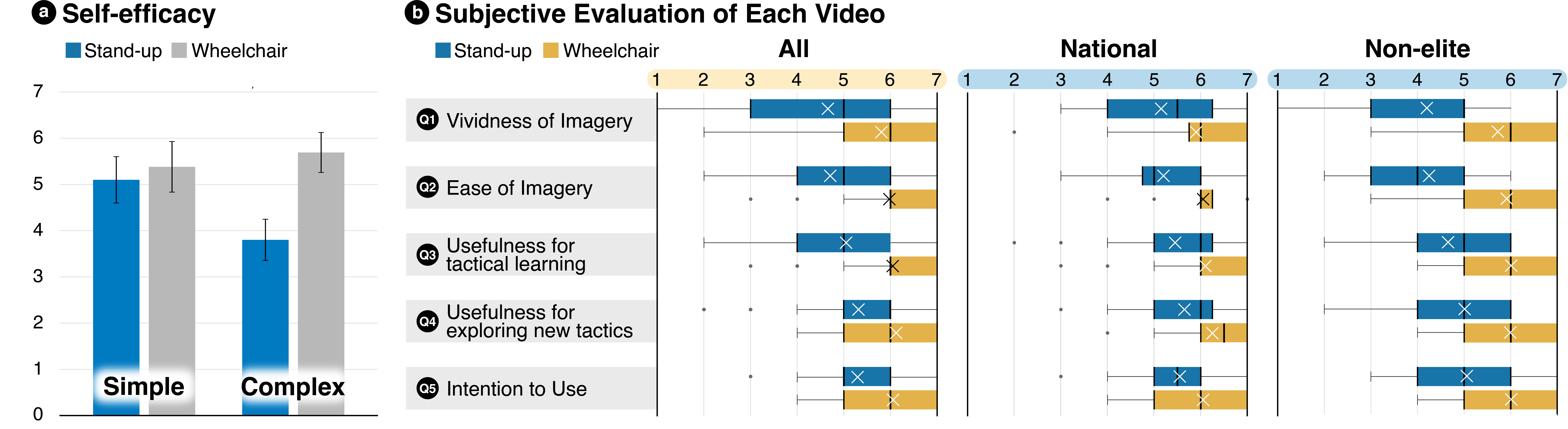}
    \caption{The results of User Study 2. (a) Participants reported significantly higher self-efficacy when watching wheelchair-converted videos compared to stand-up basketball videos. (b) For all five subjective measures, both national players and non-elite players gave higher ratings to the converted videos than to the original ones.}
    \Description{Results of User Study 2 comparing wheelchair-converted videos with stand-up basketball videos. Subfigure (a) shows significantly higher self-efficacy ratings when participants watched wheelchair-converted videos. Subfigure (b) shows that both national team players and non-elite players rated the converted videos higher than the original videos across all five subjective measures.}
    \label{fig:study2}
\end{figure*}

\subsection{Higher Self-efficacy with Reconstructed Wheelchair Basketball Videos}

Study 2 results show how reconstructed wheelchair basketball videos influenced participants’ self-efficacy, with particular attention to comparisons with stand-up basketball footage and to differences across competitive levels and tactical complexity.

As shown in Fig.~\ref{fig:study2}a, \textbf{participants reported higher self-efficacy in the W condition} compared to the S condition. Results of the ART-ANOVA revealed a significant main effect of condition 
($p = .00037$, partial $\eta^2 = .28$, $1 - \beta = 1.00$), 
and the interaction between complexity and condition was also significant 
($p = .0083$, partial $\eta^2 = .17$, $1 - \beta = 1.00$), 
indicating that the effect was particularly pronounced for complex tactics.
Additionally, a three-way interaction among group type, complexity, and condition was significant 
($p = .0317$, partial $\eta^2 = .11$, $1 - \beta = 1.00$), suggesting that the effect of reconstructed videos 
varied depending on player expertise and tactical difficulty.
\textbf{Non-elite participants consistently reported higher self-efficacy} in the W condition, both for simple tactics (W: $M = 5.57$ vs. S: $M = 4.71$) and for complex tactics (W: $M = 5.00$ vs. S: $M = 4.00$). In contrast, \textbf{national-level participants showed no difference between conditions for simple tactics} (W: $M = 5.38$ vs. S: $M = 5.62$). This suggests that they regularly use stand-up basketball footage as a learning resource and already feel confident in adapting relatively simple tactics to their own play. However, \textbf{for complex tactics, national players exhibited significantly higher self-efficacy in the W condition} (W: $M = 6.38$ vs. S: $M = 4.12$, $p < .05$, partial $\eta^2 = .11$, $1 - \beta = 1.00$). While stand-up basketball footage evoked impressions such as \textit{“not enough space to execute in a wheelchair”} or \textit{“steps and speed seem impossible to reproduce”} (P1, National; P12, Non-elite), the reconstructed videos instead elicited more positive reactions such as \textit{“with some adjustments I could actually do it”} or \textit{“I want to try it myself”} (P2, National; P19, Non-elite).

In summary, \textbf{reconstructed wheelchair basketball videos were effective in enhancing self-efficacy}. The effect was broadly evident among non-elite participants, and for national players it emerged particularly in high-complexity tactical situations, suggesting that reconstructed videos can serve as a valuable resource for supporting learning across different competitive levels.

\subsection{Consistently Higher Subjective Ratings for Reconstructed Footage}

To evaluate the effectiveness of embodiment transformation through BRIDGE, we examined participants’ subjective experiences across five perspectives in Study 2: imagery vividness, ease of imagery, usefulness for tactical learning, usefulness for exploring new tactics, and intention to use.  

As shown in Fig.~\ref{fig:study2}b,\textbf{ across all scales, participants consistently rated the reconstructed wheelchair videos significantly higher than the stand-up basketball videos}, with similar trends observed for both national and non-elite players and no significant interactions.  

For \textbf{imagery vividness} ($p = .007$, partial $\eta^2 = .17$, $1 - \beta = .99$) and \textbf{ease of imagery} ($p < .001$, partial $\eta^2 = .37$, $1 - \beta = 1.00$), participants found the reconstructed videos easier to interpret the tactics. P10 (National) explained, \textit{``The transformed videos directly connect to tactical images,''} while P12 (Non-elite) noted, \textit{``Since I have no experience with stand-up basketball, it is difficult to imagine playing in a wheelchair from those videos,''} highlighting the accessibility of the reconstructed footage.  

For \textbf{tactical learning} ($p = .013$, partial $\eta^2 = .15$, $1 - \beta = .92$) and \textbf{exploration of new tactics} ($p = .002$, partial $\eta^2 = .22$, $1 - \beta = .99$), the reconstructed videos again received higher ratings. P1 (National) remarked, \textit{``I often learn from tactic boards, but the transformed videos are overwhelmingly clearer,''} and P11 (Non-elite) similarly stated, \textit{``They make it easier to imagine details like spacing and angles.''} At the same time, some participants pointed to areas for refinement. For example, P2 (National) observed that \textit{``some movements, such as floater shots, are not yet reflected,''} while P17 (Non-elite) suggested that \textit{``the ambiguity of stand-up basketball videos can sometimes inspire new possibilities.''} These comments suggest opportunities for complementary use, a point echoed by participants who recommended combining both types of footage.  

Finally, for \textbf{intention to use} ($p < .001$, partial $\eta^2 = .32$, $1 - \beta = 1.00$), players who typically relied on stand-up basketball videos expressed that reconstructed versions would make their learning more concrete and communicable, while those who usually avoided stand-up basketball footage reported feeling more motivated if transformed videos were available. For example, P9 (National) noted, \textit{``It allows me to easily visualize the ideas I usually keep in mind, and while I used to communicate those ideas with my teammates only in words, this makes it possible to share concrete situations.''} Similarly, P18 (Non-elite) stated, \textit{``I usually do not watch stand-up basketball videos, but if they are transformed, I would feel motivated to view them more casually.''} 

Taken together, reconstructed wheelchair videos were consistently rated more favorably across all five subjective scales. Transforming footage into the wheelchair basketball context enhanced clarity, ease of processing, learning value, opportunities for tactical exploration, and intention to use. These findings indicate that \textbf{the approach can serve as a valuable design strategy for educational applications and communication with others}.

\section{Discussion}
This study introduced BRIDGE, a conversion system for wheelchair basketball, to examine how embodiment transformation affects learners’ perception, understanding, and self-efficacy. The results showed enhanced naturalness, clearer functional classification, preservation of tactical action, stronger self-efficacy, and higher subjective evaluations. These findings highlight bodily differences as valuable learning resources and suggest design principles for inclusive sports education.
In the following sections, we discuss the main findings of BRIDGE and their theoretical and practical implications.

\subsection{Embodiment Transformation for Inclusive Sports Learning}
In this section, we discuss three key findings from the user study that demonstrate the effectiveness of embodiment transformation in inclusive sports learning.

\textbf{1) Embodiment-aware Orientation Mapping Lowers Interpretation Costs in Tactical Learning.}
Accurately perceiving bodily cues is essential for tactical learning, as it enables athletes to correctly interpret the movements and intentions of players~\cite{rekik2023effect}. In this study, we proposed an embodiment-aware orientation mapping that transforms standing movements into wheelchair movements, and demonstrated that presenting trunk mobility and other bodily cues helps bridge understanding across players with different embodiments. Many participants reported that the mapped footage enabled them to discern functional classifications and appeared more natural overall. These findings suggest that aligning posture and gaze cues with the viewer’s embodied perspective reduces the cognitive effort previously required to reinterpret stand-up basketball movements within the structural constraints of wheelchair play.

However, the effectiveness of the mapping is not uniform across all situations. Some participants noted that wheelchair orientation alone was sufficient for understanding simple tactical situations, and several pointed out that distinguishing mid- and high-point players remained difficult even with embodied cues. These results indicate that while the mapping is beneficial, its impact depends on the context and the nature of the task. Future work should examine which bodily cues should be emphasized and how complementary visual representations might be integrated to optimize the design of embodied representations.

\textbf{2) BRIDGE Reduces Alienation from Embodiment Gaps and Enhances Self-Efficacy.}
In our formative study, participants reported that certain stand-up basketball movements, such as rapid accelerations or sharp directional changes in narrow spaces, were difficult to interpret within the constraints of wheelchair play. This mismatch led to a recurring sense that \textit{“the movements in the video do not correspond to my body”}, creating not only challenges for understanding but also a reduction in motivation to learn.
To address these challenges, our study showed that BRIDGE significantly enhanced participants’ self-efficacy when they viewed tactically equivalent footage reconstructed to match wheelchair-specific embodiment. Participants expressed positive reactions such as \textit{“with some adjustments I could actually do it”} and \textit{“I want to try it myself”}, indicating not only improved comprehension but also a heightened sense of actionability. These findings demonstrate that BRIDGE offers a concrete response to the long-standing call in parasport research to reinterpret and extend the meaning of sport in ways aligned with the embodied realities of athletes with disabilities~\cite{strobel2025hci}.
This also aligns with prior work in embodiment reinterpretation, which shows that while humans can flexibly reinterpret different bodily structures, such flexibility has limits: mismatched embodiments can hinder the ability to link intention and action~\cite{won2015homuncular, zhang2025becoming}.
Differences across competitive levels further highlight the robustness of this effect. For non-elite players, BRIDGE consistently bridged the gap between abstract tactical knowledge and lived bodily experience, supporting self-efficacy across a wide range of scenarios. For national-level players, its impact was especially notable in complex tactical situations, where the transformed footage strengthened confidence in attempting higher-level plays.
Taken together, these results indicate that BRIDGE mitigates feelings of alienation arising from embodiment differences and provides accessible embodied representations that enhance self-efficacy across learner groups. Importantly, this effect holds regardless of a learner’s depth of tactical expertise, suggesting that BRIDGE can function as a foundational tool for enabling inclusive engagement with shared training materials.

\textbf{3) A Foundation for Cross-Embodiment and Collaborative Learning.}
Our findings show that the transformed videos not only support individual tactical understanding but also act as practical simulations that encourage exploratory reasoning. Several participants reported that the aligned bodily context made them want to try new spacing and angles, suggesting that BRIDGE can stimulate tactical experimentation grounded in feasible movement possibilities.
At the same time, the stand-up basketball footage was found to possess distinctive value as a learning resource shaped by a different embodiment. It contains a wide range of tactical structures, movement variability, and speed dynamics that do not always emerge in wheelchair play. Some participants noted that the ambiguity of stand-up footage stimulated new possibilities, indicating that such videos retain value as a catalyst for creative thinking~\cite{wadinambiarachchi2024effects}.

However, conventional tools such as tactic boards, verbal explanations, or untransformed stand-up footage often lead to mismatches in how intentions and spatial relations are understood. In our formative study, participants pointed out that these resources made it difficult to share mental images with others, creating obstacles when using such footage for instruction or discussing tactical possibilities. In contrast, BRIDGE’s embodiment-aligned representation helped some players show what is usually only in their head, making it easier to coordinate shared understanding within the team.
Thus, stand-up footage and BRIDGE should not be viewed as mutually exclusive.
Stand-up footage serves as a resource for rich tactical structures and creative stimulation,
while BRIDGE provides a resource that aligns bodily premises and supports actionable understanding.
Together, they function in a complementary manner.

Overall, BRIDGE establishes a flexible foundation for examining tactics across different embodiments. It not only supports individual reasoning but also offers cues that facilitate team-level communication when needed. By combining the tactical richness of stand-up footage with the bodily alignment afforded by BRIDGE, learners with diverse embodiments can reconstruct tactics from multiple perspectives, moving fluidly between understanding and creative exploration.

\subsection{Beyond Wheelchair Basketball: Expanding the Reach of Embodiment Transformation}

While our study focused on wheelchair basketball as a proof of concept, the potential of conversion systems such as BRIDGE extend well beyond this domain. Their value lies in enabling the transformation of motor knowledge accumulated within one bodily configuration into forms that are accessible to learners with different bodies.

\begin{figure}[t]
    \centering
    \includegraphics[width=1.0\linewidth]{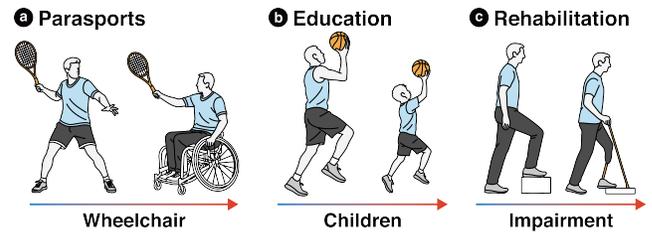}
    \caption{Potential applications of embodiment transformation. (a) Parasports: converting movements of non-disabled persons into movements of persons using a wheelchair. (b) Education: translating adult movements into child-scaled representations. (c) Rehabilitation: adapting movements to represent physical impairments.}
    \Description{Conceptual illustration showing three example application domains of embodiment transformation. The figure depicts parasports, where non-disabled movements are adapted to wheelchair-based movements; education, where adult movements are scaled to represent children; and rehabilitation, where movements are modified to reflect physical impairments.}
    \label{fig:Scenario}
\end{figure}

As illustrated in Fig.~\ref{fig:Scenario}, such systems hold potential applications not only in sports but also in diverse fields such as education and rehabilitation.
\textbf{First}, the framework of embodiment transformation is also effective in other sports domains. For instance, applying the transformation techniques developed through wheelchair basketball to sports with different bodily dynamics and ranges of motion—such as track and field, tennis, or fencing—can enable the sharing of tactical knowledge and the generation of training materials across different bodily configurations as shown in Fig.~\ref{fig:Scenario}a. By incorporating factors such as physical constraints, assistive devices, and movement styles as transformation parameters, it becomes possible to transfer tactical and technical knowledge beyond individual sports. Such applications lay the foundation for cross-sport learning, bridging differences in physical ability and competitive format.
\textbf{Second}, applications in education are equally compelling. Converting adult movements into child-scaled demonstrations can produce instructional materials that are more comprehensible to young learners as shown in Fig.~\ref{fig:Scenario}b. Educational videos are presented through adult demonstrations, which creates a substantial gap for children who must reinterpret these actions in relation to their own bodily experiences~\cite{jain2016motion, aloba2019quantifying}. Within this context, expanding resources that allow learners to project themselves into the material is essential for stimulating motivation and deepening understanding.  
Similarly, adapting examples to reflect the abilities of children with disabilities creates more inclusive learning environments, preventing bodily differences from becoming barriers to participation.  
\textbf{Third}, rehabilitation represents a natural site of application. As shown in Fig.~\ref{fig:Scenario}c, by transforming therapists’ demonstrations into movements that reflect patients’ bodily characteristics, learners can perceive actions as feasible within their own capacities. This approach does more than improve learning efficiency: it also supports motivation for continued rehabilitation and fosters psychological assurance in the recovery process. Especially in rehabilitation, where self-efficacy has been shown to be strongly associated with functional recovery and improvements in quality of life, providing support that fosters a sense of “I can do this” is crucial~\cite{gangwani2022leveraging, korpershoek2011self}. 
These application areas raise a deeper theoretical question: \textit{whose body can serve as a carrier of knowledge?} Historically, performance from non-disabled athletes has been treated as the normative reference point. However, the transformation of embodied knowledge challenges this assumption, showing that the movements and experiences of people with different bodily conditions, including those with disabilities, can also serve as valuable learning resources. In this sense, a conversion system not only provides a concrete implementation framework but also introduces a new perspective on how embodied knowledge can be understood and utilized.

In summary, embodiment transformation has the potential to extend beyond sports by broadening opportunities for learning and participation and making them more equitably accessible. Its significance for HCI lies in pointing toward design approaches that treat bodily differences not merely as constraints but as resources to be embraced and leveraged.

\subsection{Limitations and Future Work}
BRIDGE demonstrated the potential of conversion systems for supporting cross-embodiment learning, but this study has several limitations. \textbf{First}, regarding the nonsignificant results in the present analysis, the effect sizes were small and the statistical power was insufficient. This suggests that the sample size in this study may have limited the ability to detect small effects. 
Importantly, this study involved a mixed group of players with and without disabilities. While the non-disabled participants regularly trained and competed in wheelchair basketball, they do not share the competitive experience associated with being an athlete with a disability and therefore differ from players with disabilities in how they interpret embodiment mapping and self-efficacy within the context of this study. Accordingly, the present findings should not be interpreted as directly representing the learning experiences or psychological effects of wheelchair basketball players with disabilities alone.
Future research should examine long-term learning effects and practical impacts on gameplay by employing more homogeneous participant groups, as well as including participants with a wider range of disability characteristics.
\textbf{Second}, our evaluation was conducted in the specific context of wheelchair basketball. While this sport provided a rich testbed due to its combination of tactical coordination and bodily constraints, further studies are required to examine how well the approach generalizes to other sports or activities with different motor demands.
\textbf{Third}, the current system focuses on embodiment transformation through visual conversion, and multimodal feedback such as collision detection, trajectory optimization, and haptic or kinesthetic guidance has not yet been incorporated. Integrating these modalities could enable more practical tactical understanding and support exploratory learning. In particular, reproducing the tactile sensation of wheelchair contact and collision would be crucial for safely experiencing and internalizing spatial awareness and risk judgment that are unique to the sport.
\textbf{Fourth}, BRIDGE primarily aims to support perceptual learning and has not yet extended to the design of interactive training that involves physical practice or feedback. In the future, the system is expected to evolve beyond supporting players’ understanding toward providing an interactive environment where coaches can explore and refine tactics, taking bodily constraints into account and designing new plays through cross-embodiment interaction.
\textbf{Taken together}, these limitations point to several directions for future research: conducting larger-scale and longer-term evaluations with more diverse participants, developing multi-modal conversion systems, expanding applications to rehabilitation and physical education, and advancing theoretical accounts of how embodied knowledge can be transformed across bodies. Pursuing these directions will help advance the design of inclusive technologies that broaden access to learning and participation.

\section{Conclusion}

This paper introduced BRIDGE, a converter system that translates stand-up basketball demonstrations into wheelchair-based representations, and demonstrated the potential of sharing embodied knowledge across different bodily configurations. Through our study, we found that fundamental tactical actions can be effectively conveyed, while also enhancing learners' self-efficacy, increasing the vividness of their mental imagery, and accurately reflecting player classifications based on bodily functions.  
Our findings contribute to embodiment theory by clarifying how universality and specificity intersect in embodied knowledge. They also provide design principles for learning tools that preserve tactical actions, support interpretive flexibility, foster psychological resonance, and incorporate accurate representations of functional classifications.  
Beyond wheelchair basketball, we showed that converter systems hold promise for rehabilitation and education, where translating actions across bodies can broaden access to both knowledge and participation.  
Overall, this work presents a vision of cross-embodiment learning as a foundation for more inclusive technologies in HCI—technologies that not only accommodate bodily differences but also leverage them as valuable resources for design and learning.

\bibliographystyle{ACM-Reference-Format}
\bibliography{sections/references}

\end{document}